\documentclass{article}
\usepackage{caption}
\usepackage[utf8]{inputenc}
\usepackage{amsmath,amsthm,bm,mathrsfs,mathtools,hyperref,lipsum}
\usepackage[numbered]{bookmark}
\usepackage{mdframed}
\usepackage{courier}
\usepackage{authblk}
\hypersetup{
    colorlinks,
    citecolor=black,
    filecolor=black,
    linkcolor=black,
    urlcolor=black
}
\usepackage[a4paper, total={7in, 10in}]{geometry}
\usepackage{graphicx, float, multicol, balance}
\usepackage[version=4]{mhchem}
\usepackage[sorting=none]{biblatex}
\title{Origin of Biological Homochirality by Crystallization of an RNA Precursor on a Magnetic Surface}

\author[1$\dagger$]{S. Furkan Ozturk}
\author[2]{Ziwei Liu}
\author[2]{John D. Sutherland}
\author[3]{Dimitar D. Sasselov} 

\affil[1]{Department of Physics, Harvard University, Cambridge, MA 02138, USA}
\affil[2]{MRC Laboratory of Molecular Biology, Cambridge Biomedical Campus, Cambridge CB2 0QH, UK}
\affil[3]{Department of Astronomy, Harvard University, Cambridge, MA 02138, USA}
\affil[$\dagger$]{\href{mailto:sukrufurkanozturk@g.harvard.edu}{\normalfont\texttt{sukrufurkanozturk@g.harvard.edu}}}

\date{\today}

\begin{document}
\maketitle

\begin{abstract}
    Homochirality is a signature of life on Earth yet its origins remain an unsolved puzzle. Achieving homochirality is essential for a high-yielding prebiotic network capable of producing functional polymers like ribonucleic acid (RNA) and peptides. However, a prebiotically plausible and robust mechanism to reach homochirality has not been shown to this date. The chiral-induced spin selectivity (CISS) effect has established a strong coupling between electron spin and molecular chirality and this coupling paves the way for breaking the chiral molecular symmetry by spin-selective processes. Magnetic surfaces can act as chiral agents due to the CISS effect and they can be templates for the enantioselective crystallization of chiral molecules. Here we studied the spin-selective crystallization of racemic ribo-aminooxazoline (RAO), an RNA precursor, on magnetite (Fe\textsubscript{3}O\textsubscript{4}) surfaces, achieving an unprecedented enantiomeric excess of about 60$\%$. Following the initial enrichment, we then obtained homochiral crystals of RAO after a subsequent crystallization. Our work combines two necessary features for reaching homochirality: chiral symmetry-breaking induced by the magnetic surface and self-amplification by conglomerate crystallization of RAO. Our results demonstrate a prebiotically plausible way of achieving systems level homochirality from completely racemic starting materials.
\end{abstract}

\begin{multicols}{2}

Understanding the origins of biomolecular homochirality is essential for understanding the origins of life, and the origin of homochirality remains a long-standing mystery since Pasteur discovered the molecular asymmetry of organic compounds in 1848 \cite{pasteur1848relations}. Achieving a homochiral state early in the prebiotic synthesis of the monomers would be very beneficial for successful polymerization and the overall robustness of the entire synthetic network \cite{joyce1984chiral, blackmond2010origin}. The prebiotic need for high yields combined with high selectivity requires a persistent and well-matched pair of a chiral symmetry-breaking agent and an amplification mechanism.

One of the central molecules for prebiotic synthetic networks emerged in 1970 from the work of Sanchez and Orgel \cite{sanchez1970studies}, who identified the aminooxazolines (ribo, arabino, xylo and lyxo) as novel organic intermediates useful for the synthesis of a wide variety of nucleotides, the monomers of nucleic acids (RNA and DNA). Follow-up studies \cite{springsteen2004selective, anastasi2006direct} culminated in the prebiotic synthesis of pyrimidine nucleotides \cite{powner2009synthesis} with aminooxazolines as the precursors, and the consequent development of an entire synthetic network for nucleotides and amino acids by the Sutherland lab \cite{patel2015common}. Blackmond and coworkers emphasized ribo-aminooxazoline (RAO)'s centrality in solving the origin of homochirality due to its crystallization properties \cite{hein2011route} but the search for a prebiotically plausible mechanism that can induce and amplify a chiral bias has remained to be an open problem.

Biological systems comprise many homochiral molecules. In principle such a state could be achieved by separately resolving each individual chiral compound. However, a more attractive solution would be to establish the homochirality in a compound from which propagation of homochirality to the whole system could occur. Emergence of homochirality at the stage of RAO sets the stage for the propagation of homochirality through RNA to peptides and thence through enantioselective catalysis to metabolites.

%This propagation avoids the requirement for separate solutions to the homochirality problem for all chiral molecules in the biological system.

In our recent work, we proposed a symmetry-breaking agent that can trigger such an amplification in the reduction reactions leading up to the production of the chiral 3-carbon sugar glyceraldahyde (itself the precursor to RAO). Our proposed mechanism employs the strong coupling between the electron spin and molecular chirality as established by the chiral induced spin selectivity (CISS) effect \cite{ozturk2022homochirality}. We identified evaporative lakes with authigenic iron-oxide sediments as prebiotically plausible settings \cite{sasselov2020origin} in which enantioselective reduction reactions can occur close to the magnetized surface by spin-polarized photo-electrons released from the magnetite by solar UV light.

In this work, we demonstrate that the spin-polarized surfaces themselves are chiral agents breaking the chiral symmetry at the surface, in a further expansion of the role of the CISS effect, and the simultaneous conglomerate crystallization provides the necessary amplification to reach a homochiral state (Figure 1). Here we report the spin-selective homochiral crystallization of the ribonucleotide precursor, ribo-aminooxazoline (RAO), on magnetite surfaces, from its completely racemic solution (Figure 2).

\pdfbookmark[section]{Homochirality of RNA and the Central Role of RAO}{sec1}
\section*{Homochirality of RNA and the Central Role of RAO}

RNA is thought to have played two major roles in the origin of life. The sequence of the nucleobases attached to the sugar phosphate backbone constitutes genetic information which can be passed from generation to generation by replication via a complementary strand through Watson-Crick base pairing. The sequence of an RNA also dictates its shape, and it is the adoption of a wide variety of shapes that endows RNA with its catalytic ability. Enantiomeric purity of the nucleotide components of RNA is crucial to both its roles. Replication via a complementary strand proceeds by way of an A-form duplex and this is not possible if the nucleotide components deviate significantly from enantiomeric purity. Incorporation of a nucleotide of opposite handedness into an RNA strand changes its shape and this might endow a new or improved catalytic ability, but the chirality switch cannot be passed to subsequent generations by replication, so the potentially beneficial change is non-hereditable. Establishing a mechanism whereby the nucleotide building blocks of RNA might have been synthesized enantiomerically pure is thus crucial to understanding how Earth’s RNA-based life originated. Peptides too depend on the enantiomeric purity of their component amino acids. In extant biology, the correlation of D-ribonucleotides with L-amino acids is established in two principal ways. Ribonucleotides and chiral amino acids are biosynthesized with high stereoselectivity and L-amino acids are attached to transfer RNA (tRNA) composed of D-ribonucleotides by stereoselective aminoacyl-tRNA synthetases. The idiosyncratic behavior of amino acids makes the synthesis of all L-amino acids under early Earth conditions a daunting challenge, but recent findings suggest that this hurdle might be side-stepped. Prebiotically plausible chemical means of attaching amino acids to tRNA analogs have been discovered and proceed with high-level control of relative stereochemistry. Thus, for example, the L- over D- stereoselectivity for attachment of alanine to a tRNA acceptor stem mimic composed of D-ribonucleotides is of the order of 10:1 \cite{wu2021interstrand}. This suggests that the overall homochirality problem of prebiotic chemistry might be reduced to the problem of making ribonucleotides in enantiopure form\textemdash racemic amino acids might suffice.

RAO emerged as a potentially important ribonucleotide precursor over 50 years ago, but there remained several issues to be dealt with before its true potential was realized. Sanchez and Orgel showed that RAO (which they synthesized from ribose and cyanamide) was a highly crystalline compound which underwent reaction with cyanoacetylene to generate $\alpha$-cytidine, but subsequent conversion to the natural $\beta$-anomer was very inefficient \cite{sanchez1970studies}. Furthermore, the synthesis of ribose is one of the persistent challenges of prebiotic chemistry – even though the formose reaction is now very well understood, it still cannot be controlled to produce more than a trace of ribose as part of a complex mixture. Nevertheless, the attractiveness of RAO prompted further research and Joyce showed that it could be sequestered from simpler mixtures of sugars than those made by the formose reaction by reaction with cyanamide followed by crystallization \cite{springsteen2004selective}. Joyce also reported more on the crystallization behavior of RAO and showed that it crystallized in the chiral $P 2_1 2_1 2_1$ space group and that crystals grown from solutions of \textit{rac}-RAO were composed of twinned clusters with individually homochiral domains. The problem with the provenance of ribose was solved when Sutherland showed RAO could be obtained, along with lesser amounts of the other pentose aminooxazolines, by reaction of glyceraldehyde with 2-aminooxazole, the latter itself deriving from the reaction of glycolaldehyde with cyanamide \cite{anastasi2006direct}. The same group also found that non-racemic solutions of RAO gave rise to crystals with increased enantiomeric excesses (ee’s) and above a threshold ee, enantiopure crystals resulted. This behavior is consistent with a conglomerate susceptible to twinning. The conglomerate nature of RAO was finally proven by Powner \cite{islam2017prebiotic}. Then high-yielding prebiotic syntheses of ribonucleotides from aminooxazolines were reported by Sutherland. First, it was shown that arabino-aminooxazoline could be elaborated to pyrimidine nucleoside cyclic phosphates \cite{powner2009synthesis}, then it was shown that RAO could be converted to pyrimidine nucleotides by a reaction sequence involving a photochemical inversion of the anomeric stereocentre \cite{xu2017prebiotically}. In the meantime, Blackmond had shown that enantiomerically-enriched proline could participate in the reaction of 2-AO with glyceraldehyde and give enantiopure RAO after crystallization \cite{hein2011route}. {\it What has remained elusive however is how racemic compounds could give rise to enantiopure RAO purely by means of a process controlled only by the environment.}

\end{multicols}

\begin{figure}[H]
    \centering
    \includegraphics[width=\textwidth]{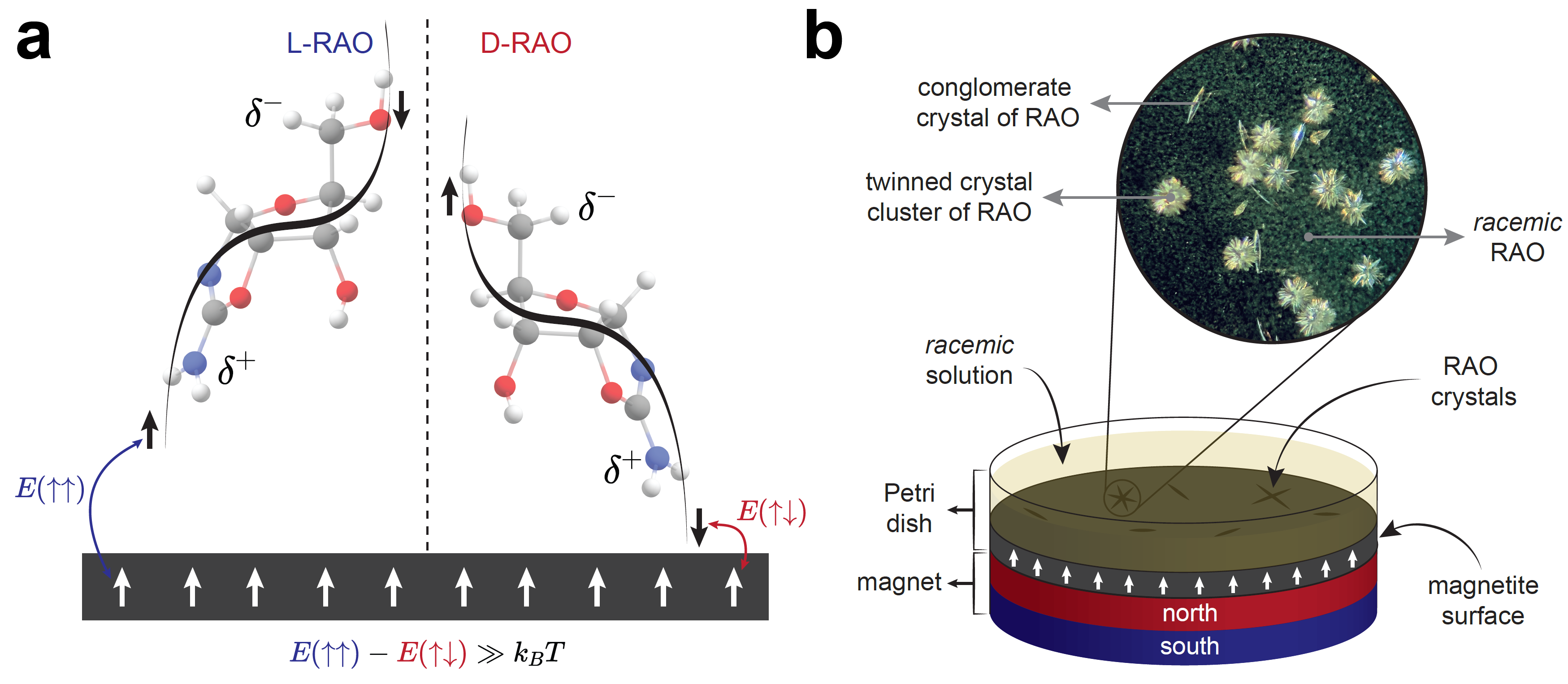}
    \caption{\textbf{The mechanism of spin-selective crystallization due to the CISS effect and the experimental setup.} \textbf{a.} As molecules approach a surface, they transiently acquire an induced charge polarization. Due to the CISS effect, transient charge polarization of a chiral molecule is accompanied by spin polarization. The spin state associated with the charge poles is determined by the handedness of the chiral molecule. Because the magnetic surface itself is spin polarized it kinetically favors (akin to a seed crystal) the enantiomer whose transient spin state results in a lower energy spin-exchange interaction. The lower energy overlap with the magnetic surface is singlet-like (red, $\uparrow\downarrow$) and the higher energy overlap is triplet-like (blue, $\uparrow\uparrow$). The energy difference between these two configurations is higher than the room temperature, $k_BT$, therefore the effect robustly manifests itself. \textbf{b.} Schematic of the setup used in the crystallization experiments and a sample microscope image of the RAO crystals on a magnetite surface from a direct crystallization experiment. The image shows the magnetite surface as the black background and the needle-shaped conglomerate crystals of RAO formed on the surface, as well as the twinned crystals with stochastically arranged needles of D and L-RAO, and racemic RAO in the form of a flaky powder suspended in the water column above the surface.}
    \label{Fig.1}
\end{figure}

\begin{multicols}{2}

\pdfbookmark[section]{Chiral Induced Spin Selectivity and Spin Selective Chemistry}{sec2}
\section*{Chiral Induced Spin Selectivity and Spin Selective Chemistry}

The CISS effect has established a robust coupling of electron spin to molecular chirality. The initial experiments have shown that the electron transfer through a chiral monolayer is spin-dependent and the preferentially transferred spin state depends on the handedness of the monolayer \cite{ray1999asymmetric, gohler2011spin}. The experiments have achieved near-perfect spin filtering at room temperature\textemdash showing the robustness of the coupling \cite{naaman2012chiral}. Later work has proved that the coupling established by the CISS effect can be manifested in various ways from spintronic applications to long-range electron transfer in biology \cite{naaman2022chiral, michaeli2016electron, naaman2015spintronics}. More recent studies have shown that spin selective behavior exists for freely diffusing molecules near magnetic surfaces \cite{metzger2020electron, bloom2020asymmetric}, and even for achiral reagents  \cite{metzger2021dynamic, bhowmick2022spin}. These studies have motivated us to consider the spin-selective processes near magnetic surfaces due to the CISS effect to impose a chiral bias on prebiotic chemistry.

Recently, we proposed closed-basin evaporative lakes with authigenic ferrimagnetic sediments (e.g. magnetite, greigite) as plausible prebiotic environments in which enantiospecific processes can be carried out due to the CISS effect \cite{ozturk2022homochirality}. Our mechanism employed UV-ejected photo-electrons from magnetite surfaces as chiral agents in reductive synthesis because of the helical character of such electrons in the close vicinity of the magnetic surface (Fig. 1. in \cite{ozturk2022homochirality}). We proposed that these helical electrons can achieve kinetic resolution in the reduction of chiral molecules for which the reaction rates for enantiomers differ by $\exp\left(\frac{2H_{\text{SO}}}{k_BT}\right)$. Here, $H_{\text{SO}}$ is defined as the effective coupling due to a combination of spin-orbit and spin exchange interactions of a chiral molecule with an electron, $k_B$ is the Boltzmann constant, and $T$ is the temperature (Fig. 2. in \cite{ozturk2022homochirality}). As a suitable chemistry to apply our idea, we proposed cyanosulfidic chemistry which uses photo-ejected hydrated electrons to drive the reductive synthesis of sugars \cite{patel2015common, xu2018photochemical}. In particular, we focused our attention on the spin-selective reduction of the cyanohydrin of glycolaldehyde that is producing glyceraldehyde, the first chiral sugar, and an RNA precursor\textemdash after the hydrolysis of the resultant imine. Although this proposed reduction is still a worthy experiment, the isomerization of glyceraldehyde to dihydroxyacetone and the cyanide liberating equilibrium of the reagent cyanohydrin, make it a harder reaction to study. Moreover, freely diffusing molecules with a single chiral center are likely to display less enantioselective behavior due to the reduced coupling of the electron spin to the molecular frame. Therefore, in this work, we studied the crystallization of the stable RNA precursor, RAO, as a static process with higher selectivity, on a magnetic surface. 

We should emphasize that, if realized, enantioselective synthesis of glyceraldehyde reinforces the spin-selective crystallization of RAO as glyceraldehyde's chirality directly determines the chirality of RAO as can be seen in the blue box in Figure 3b. However, the crystallization process does not rely on starting from an enantioenriched solution of RAO, therefore our results presented here are self-sufficient to achieve a homochiral RNA world.

\pdfbookmark[section]{Magnetic Surfaces as Chiral Agents}{sec3}
\section*{Magnetic Surfaces as Chiral Agents}

Magnetic surfaces themselves can be chiral agents due to the CISS effect and surface processes such as crystallization or adsorption on a magnetic surface can be enantioselective. As demonstrated by the early CISS experiments, electron flow from a surface through a chiral monolayer is spin-dependent. The same effect manifests itself when a chiral molecule acquires an induced charge polarization because the latter is nothing but a transient electron flow inside a chiral potential. And due to the CISS effect, charge polarization is accompanied by spin polarization. This transient, induced charge polarization can be due to intermolecular interactions among chiral molecules or between a chiral molecule and a surface \cite{michaeli2016electron, naaman2019chiral}. 

When a chiral molecule approaches a surface, the electron density of the molecule re-distributes itself which gives rise to an induced charge dipole. And for a chiral molecule, this transient electron flow is spin-dependent due to the CISS effect and gives rise to a transient spin polarization as shown in Figure 1a. This spin polarization is realized along the chiral molecular axis and it is dependent on the handedness of the chiral molecule. In Figure 1a, the right-handed (D) enantiomer has the minority spin ($\downarrow$) on its positively charged pole ($\delta^+$), whereas the left-handed (L) one has the majority spin ($\uparrow$). This is how the chiral symmetry is broken by a magnetic surface which is itself spin polarized: the spin exchange interaction between the surface electron spins and the transiently spin-polarized chiral molecules is higher or lower depending on the handedness of the molecule. The favorable, lower energy interaction with the surface corresponds to a singlet-like ($\uparrow\downarrow$), and the penalized, higher energy interaction corresponds to a triplet-like ($\uparrow\uparrow$) overlap of the spins. Therefore, a spin-polarized surface kinetically traps an enantiomer based on its spin polarization and breaks the chiral symmetry.

Enantioselective behavior on magnetic surfaces due to the CISS effect has been demonstrated with the adsorption of chiral molecules like double-strand DNA and L-cysteine on magnetized ferromagnetic films \cite{banerjee2018separation}. Moreover, enantioseparation on magnetic surfaces by conglomerate crystallization has been realized with several amino acids with high efficiency \cite{tassinari2019enantioseparation, bhowmick2021simultaneous}. Especially, these studies lay the groundwork for the presented results. We apply the same idea using a different molecule on a different magnetic substrate relevant to the prebiotic conditions. And similarly, we use the magnetic surface analogous to a crystal seed, breaking the chiral symmetry by promoting the crystallization of one enantiomer. It should be noted that this is a kinetic-entrainment-like phenomenon and not a thermodynamic effect. Therefore, if all molecules are allowed to crystallize the selective effect cannot be observed.

Naturally, the CISS-driven interaction between chiral molecules and magnetic surfaces works both ways around: a chiral molecule can also selectively spin polarize a magnetic surface and the direction of this polarization is determined by the handedness of the chiral molecule. This effect has also been demonstrated by the chirally selective magnetization switching in a ferromagnet upon the adsorption of chiral molecules \cite{ben2017magnetization}. In that work, Ben-Dor \textit{et al.} have shown that chiral molecules can robustly affect the magnetization of surfaces due to the long-range spin-exchange interaction. Chirality-induced magnetization switching combined with the presented results pave the way for positive feedback between chiral molecules and magnetic surfaces which can purify the magnetization of the iron-oxide sediments and thus promote the homochiral crystallization of RAO.

\pdfbookmark[section]{Feedback Between the Magnetic Surface and RAO}{sec4}
\section*{Feedback Between the Magnetic Surface and RAO}

Utilizing the CISS-based interaction between magnetic surfaces and chiral molecules, it is possible to envision a positive feedback loop between the magnetite surfaces and RAO in a prebiotic setting as shown in Figure 3b. In a wide range of natural depositional environments, authigenic magnetite sediments form and magnetize under the Earth’s field, resulting in a chemical remanent magnetization (CRM) \cite{dunlop2001rock}. Sediments that acquired a chemical remanence after the most recent geomagnetic reversal are expected to have a statistically uniform remnant magnetization on a hemisphere scale within a shallow depth interval \cite{karlin1987authigenic}. This initial natural bias of the magnetic domains can break the chiral symmetry if racemic RAO crystallizes on the magnetic surface. However, due to the imperfect alignment of the magnetic domains, the first crystallization attempt can only induce a small enantiomeric imbalance. However, after the enriched crystals are dissolved due to a sporadic water flow and re-crystallize on the magnetic surface again, the selectivity will be higher due to the asymmetric crystallization of RAO \cite{anastasi2006direct} and further magnetic seeding due to the CISS effect (See Extended Fig. 1a). When these nearly pure RAO conglomerates grow on the surface, they can simultaneously interact with the surface spins and switch the magnetization of the surface. This process can purify the magnetic domains along the chiral molecular axis of RAO. And because these sedimentary rocks are no longer small superparamagnetic particles, this \textit{CISS induced magnetic diagenesis} can permanently lock the surface spins along one direction, unless a large (much larger than Earth's field) coercive field is applied. When the surface spins are locked once and for all, they can induce efficient enantioselectivity for the upcoming crystallizations and the magnetized surface can separate homochiral RAO crystals in just a few cycles. 

This feedback mechanism can be more effectively realized over a small area such as the scouring zone of an incoming stream as shown in Figure 3a. At the scouring zone, the authigenic magnetic sediments are exposed, wet-dry cycles are more frequently realized (like a shore), and an incoming stream can constantly feed the surface with racemic RAO. With the described feedback mechanism, magnetized magnetite sediments at the scouring zone can filter out homochiral RAO crystals from the incoming racemic stream. 

\end{multicols}

\begin{figure}[h]
    \centering
    \includegraphics[width=\textwidth]{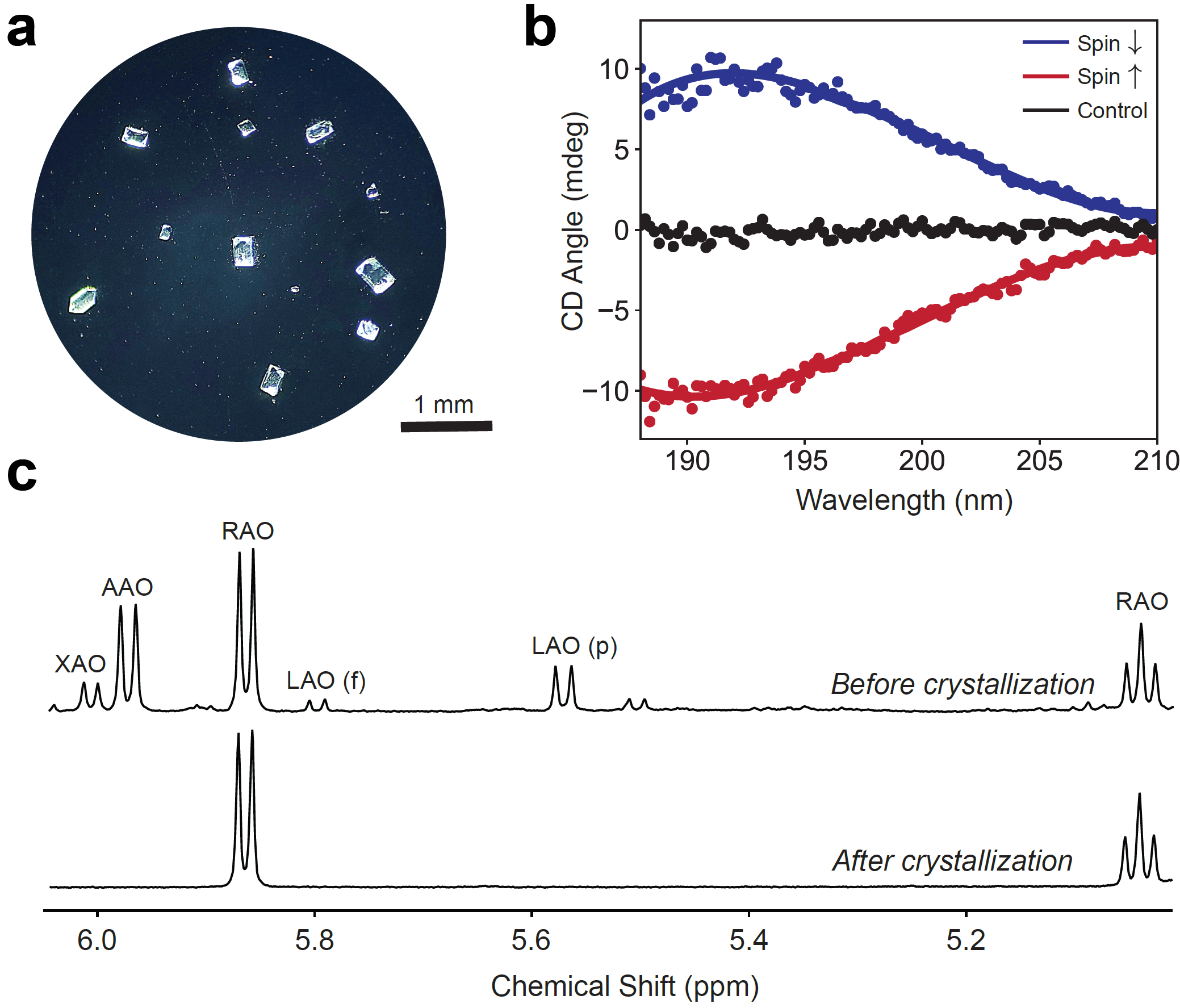}
    \caption{\textbf{Stereoselective and enantioselective crystallization of RAO.} \textbf{a.} A microscope image of the nearly enantiopure RAO crystals formed on a magnetite surface from their racemic solution. \textbf{b.} Circular dichroism spectra of the crystals formed on the magnetite surface. The red (blue) spectrum corresponds to D-RAO (L-RAO) crystals formed when the magnetite surface is magnetized parallel (anti-parallel) to the surface normal. Both spectra are obtained when racemic RAO is re-crystallized and they show a cumulative ee of about $60\%$ for the whole surface. The black spectrum corresponds to crystallization on a non-magnetic, silicon surface in the presence of a magnetic field and it shows no selectivity. The latter control experiment shows that the enantioselective effect is due to the spin-exchange interaction, not due to the applied magnetic field. \textbf{c.} \textsuperscript{1}H NMR spectrum (400 MHz, H\textsubscript{2}O/D\textsubscript{2}O 90:10) before and after the direct crystallization of RAO. Before the crystallization, the solution is a mixture of four aminoxazolines: RAO/AAO/LAO(f for furanose, p for pyranose)/XAO (42:30:18:10). After the crystallization, the redissolved crystals only contain RAO within the limits of \textsuperscript{1}H NMR detection. Thereby, mere crystallization on the magnetic surface stereoselectively and enantioselectively purifies RAO.}
    \label{Fig.2}
\end{figure}

\begin{multicols}{2}

\pdfbookmark[section]{Flow Chemistry}{sec5}
\section*{Flow Chemistry}

The mechanism we have shown here benefits from the presence of flow and streams as previously considered by Ritson \textit{et al}.\cite{ritson2018mimicking}. As such, we conceive of the enantioselective crystallization of RAO on the scouring zone of an evaporative lake by an incoming stream carrying the racemic solution of sugar aminooxazolines. In this scenario, the crystallization takes place on the shallow shores of the magnetite lake which can undergo multiple wet-dry cycles allowing several crystallization-dissolution cycles of RAO. The flow also scours the soft, muddy material off the surface and exposes the authigenic magnetite sediments on which enantioselective processes can take place.

Moreover, while the flow feeds the magnetic surface with a racemic solution of RAO it simultaneously washes the other enantiomer in the solution and the racemic powder suspended in the bulk away from the zone and allows for higher enantioselectivity. 

The presence of flow also keeps the area around the crystal surface locally racemic such that the depletion of one enantiomer does not lead to an increase in the concentration and subsequent crystallization of the other. Therefore, the flow maintains the enantioselective crystallization out of equilibrium, preventing the system from reaching thermodynamic equilibrium. This is crucial, as the enantioseparation mechanism we have shown is a kinetic effect and the system would reach its racemic equilibrium if the solution is not maintained racemic.

Finally, it is more convenient to conceive of the feedback loop we propose on a smaller region, rather than the whole surface of the lakebed. With selective crystallization and magnetization cycles, the spin purity of the magnetic surface around the scouring zone can be locked with less material, and this small bottleneck region over which the material flows can act as a chiral filter for the incoming RAO. Nevertheless, achieving our enantioseparation mechanism is not a fine-tuned geochemical scenario and one can come up with other (or more refined) prebiotically plausible scenarios compatible with this process. We, therefore, emphasize the robustness and the strength of our mechanism rather than the specific geochemical scenario that can accommodate the enantioseperation process.

\end{multicols}

\begin{figure}[h]
    \includegraphics[width=\textwidth]{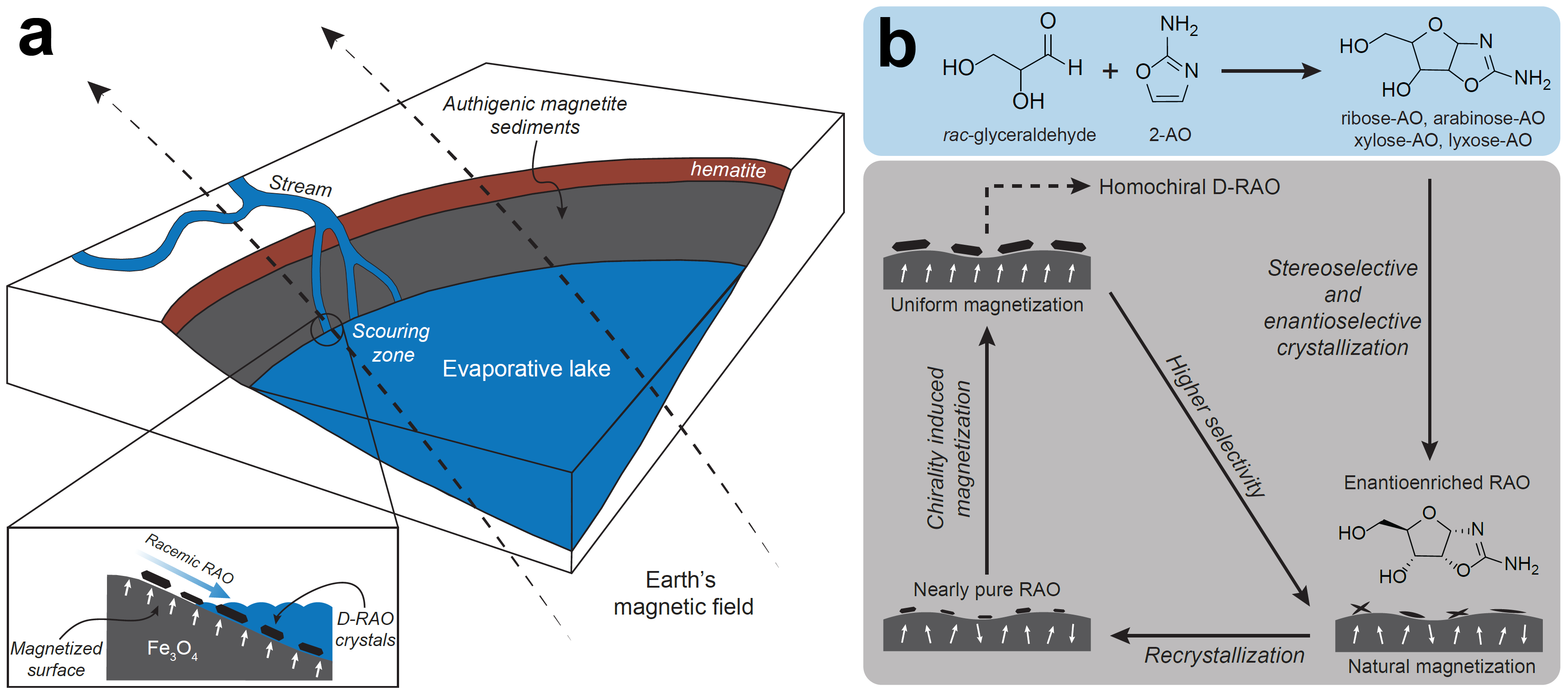}
    \caption{\textbf{An evaporative lake with magnetic sediments can accommodate spin-selective processes between an RNA precursor and a magnetized surface.} \textbf{a.} An evaporative lake contains authigenic magnetite sediments magnetized by the Earth's magnetic field. An incoming stream with racemic aminooxazolines scours the muddy material off the surface. As the lake evaporates D-RAO selectively crystallizes on the magnetic surface. \textbf{b.} An incoming stream carries the reaction products of racemic glyceraldehyde and 2-aminooxazole: racemic RAO, AAO, XAO, and LAO. Initially, the magnetic sediments are authigenically magnetized and carry a bias along the Earth's field. This natural bias enables the enantioselective crystallization of RAO which also sterically purifies RAO from the mixture of aminooxazolines. After a small enantiomeric imbalance is induced, sporadic water flow redissolves the RAO crystals and in a subsequent dry phase, RAO recrystallizes on the surface with higher ee. As these nearly pure conglomerate crystals accumulate more material, they cover a larger area of the surface and flip the spin of magnetic domains along the chiral molecular axis. This chirality-induced magnetization process increases the magnetization of the surface and allows for a positive feedback loop between the magnetic surface and the chiral crystals. As a result, the surface with higher magnetization induces higher ee for the upcoming crystallizations, and eventually, enantiopure D-RAO is obtained.}
    \label{Fig.3}
\end{figure}

\begin{multicols}{2}

%Results
\pdfbookmark[section]{Enantioselective Crystallization of RAO on Magnetite}{sec6}
\section*{Enantioselective Crystallization of RAO on Magnetite}

We have studied the crystallization of RAO from its racemic solution on magnetite surfaces. We fabricated the magnetite surfaces as thin films (200 nm) on silicon substrates following the procedure by Jubb and Allen \cite{jubb2010vibrational} (Supplementary Information Section 5). We used electron beam evaporation to deposit 100 nm iron on Si (100) surface and then heated the sample at 175$^{\circ}$C for 4 hours to promote the oxidation of iron, producing a magnetite film of about 200 nm. We characterized the samples by Fourier-transform infrared spectroscopy (FTIR) and confirmed the complete conversion of iron to magnetite (Figure S5). We further analyzed the surface roughness and magnetic properties of the samples by atomic force microscopy (AFM) and superconducting quantum interference device (SQUID) (Supplementary Information Section 6). We placed the magnetite surfaces horizontally in a Petri dish and placed a magnet just below the surface as shown in Figure 1b. We used the magnet to spin-polarize the magnetite surface and measured the magnetic field to be 325 mT at the surface location. It is important to place the magnet such that the magnetization direction is parallel to the surface normal to maximize the helical character of the surface electrons.

The \textsuperscript{1}H NMR spectra confirmed the stereoselective crystallization of RAO, as has been previously reported \cite{springsteen2004selective, anastasi2006direct}. Having identified the composition of the crystals by NMR, we measured the chiro-optical properties by circular dichroism (CD) spectroscopy. We obtained a non-zero CD signal for the individual needle-shaped crystals of RAO, however, the rosette-shaped crystals did not consistently give a CD signal. We further analyzed the rosette-shaped crystals by X-ray crystallography and found that the individual arms of the rosettes contain only one enantiomer yet the crystal as a whole is racemic. Therefore, we found that the rosettes form due to the twinning of needle-shaped homochiral domains. When we analyzed rosettes with less homochiral domains we obtained a non-zero CD signal with a randomly varying sign, however, the rosettes with more branching did not give a CD signal. We also did a control experiment by crystallizing the enantiopure compound and still observed the formation of rosette-shaped crystals. Therefore, we concluded that the twinned, rosette-shaped crystals form due to a stochastic arrangement of homochiral domains. Because this stochastic twinning occurs on an existing crystal face, it nullifies the enantioselection seeded by the magnetic surface. For the experiments in which we primarily obtained needles on the surface (Figure S11), we observed a surface-wide ee (in addition to the ee of the individual crystals), however, for most of the cases, we could not consistently avoid the formation of the twinned crystals. Therefore, we did not observe consistent selectivity for direct crystallization experiments on magnetite surfaces. However, if the rosette formation can be avoided by the change of conditions, such as concentration, buffer, or pH, it should be possible to observe selectivity for direct crystallization.

We proceeded with the re-crystallization of RAO. We synthesized racemic RAO by the reaction of D- and L-ribose with cyanamide on a large scale. We confirmed that synthesized RAO is fully racemic with a CD measurement before the crystallization experiments (Figure S17). We then prepared a 65 mM solution of the racemic RAO in pure water and placed it on the magnetized magnetite surface. We followed the same crystallization procedure described above and obtained diamond-shaped crystals on the magnetite surface as shown in Figure 2a. We collected the crystals with tweezers, dissolved them in pure water, and obtained their CD spectra. First individually and after for the whole surface. We confirmed that diamond-shaped crystals of RAO are individually homochiral by CD and X-ray diffraction measurements. These diamond-shaped crystals and the individual needle-shaped ones both gave identical X-ray diffraction patterns and they all belong to the $P2_1 2_1 2_1$ chiral space group. However, with the diamond-shaped crystals, we did not observe twinning when the crystallization is stopped soon after the crystals are observed. When we waited longer, as the crystals grew in size we observed that a crystal face became a seed for the other enantiomer. However, these twinned crystals mostly contained two to three macroscopic, homochiral domains unlike the rosettes containing many arms, for which each arm is an individual, homochiral domain. Also, for the re-crystallization experiments, twinned diamond crystals appeared well after the individual diamonds were formed and visible, therefore we could stop the experiment before the twinning started to take over. However, for the direct crystallization experiments, rosettes appeared almost simultaneously with the needles so there was no slow progression of the twinning we could control.

Having confirmed that diamond-shaped crystals are individually homochiral, we dissolved all the crystals on the magnetite surface together in water and obtained the CD spectrum for the whole surface. To our delight, not only the individual crystals showed optical activity but also the entire surface did. As the smoking gun of the CISS-driven phenomenon, when we reversed the magnetic field direction (therefore the spin state of the surface electrons) the observed CD signal reversed in sign. As shown in Figure 2b, left-handed RAO crystals (blue curve) dominated the down-spin ($\downarrow$, south pole) surface and the right-handed ones dominated the up-spin ($\uparrow$, north pole) surface. \textit{We obtained an ee of about $60\%$ for the entire surface, for both spin directions starting from a completely racemic solution of RAO.} For this step, it was crucial to collect the crystals early, as the enantioselectivity of the magnetic surface is a kinetic entrainment effect and beyond a certain point, ee decreased with the increasing crystallization yields and due to enantiomorphous twinning. Crystals were usually visible after several hours and we collected them after about 6 hours to a day. We collected 48 and 16 individual crystals for the up-spin and down-spin experiments respectively for the data shown in Figure 2b. Due to the high number of crystals and the repeatability of the results (Extended Data Fig. 1b), we can rule out the presence of a net ee due to statistical fluctuations of homochiral crystals.

We investigated the reason why re-crystallization of RAO gives crystals with different morphology (diamond-shaped) compared to direct crystallization of the reaction products of glyceraldehyde and 2-aminooxazole (needle-shaped) and crystallized RAO with added aminooxazolines (AAO, LAO, and XAO) one by one. We found that in the presence of XAO, RAO crystallizes as needle-shaped crystals whereas AAO and LAO do not visibly modify the crystallization habit of RAO. Moreover, we observed a concentration-dependent formation of the rosette-shaped twinned crystals only when RAO is crystallized in the presence of XAO (Extended Data Fig. 2a). At higher concentrations, when XAO is present in the solution, RAO forms rosette-shaped crystals and the crystal morphology is altered by the relative amount of XAO (Extended Data Fig. 2c), therefore we conclude that XAO is a crystal habit modifier for RAO (See Scheme 1 in \cite{weissbuch1995understanding}). We also found that at higher relative concentrations (e.g. 1:1) XAO is embedded in the crystal lattice structure of RAO as an impurity with $4.8(4)\%$ abundance, as detected by X-ray diffraction analysis (Figure S29). Finally, we found that XAO affected the RAO crystallization only when the relative stereochemistries matched (e.g. D-RAO with D-XAO)\textemdash we did not observe any effect of XAO when we crystallized L-RAO with D-XAO. These findings show the necessity of re-crystallization in order to observe enantioenrichment to enantiopurity, as by removing XAO from the solution racemizing stochastic twinning is circumvented.

As a control experiment, we recrystallized RAO on a non-magnetic, silicon, surface in the presence of a magnetic field of the same strength and we collected the crystals from the non-magnetic surface and obtained the black spectra in Figure 2b. As seen, the surface with no net spin polarization does not induce any enantioselectivity in the presence of a magnetic field, as shown by Pasteur \cite{pasteur1848relations}. The control experiment confirms that the observed selectivity is not due to the magnetic field but due to the spin-exchange interaction. The control experiments on the non-magnetic surface and the consistent flipping of the ee (See Extended Data Fig. 1b) with the flipping magnetic pole direction ensure that the obtained ee is physical and not due to the contamination of surfaces with chiral impurities. 

We repeated the re-crystallization experiment on magnetite for both pole directions multiple times and found that on average we can get around $35\%$ ee for the entire magnetite surface. We then considered this value as a typical outcome from a racemic solution and then crystallized the enriched crystals on the magnetic surface one more time. We found that above a starting ee of about $25\%$ we can obtain completely enantiopure crystals (See Extended Data Fig. 1a). Therefore, in just two crystallization steps on the magnetic surface, we could achieve homochirality from a completely racemic mixture.

\pdfbookmark[section]{Discussion}{sec7}
\section*{Discussion}

We have demonstrated an efficient mechanism to resolve racemic RAO, an important RNA precursor, on a prebiotically available mineral surface, magnetite $\ce{Fe3O4}$, and obtained homochiral crystals of RAO in two crystallization steps.

Our mechanism features the two requisites to reach homochirality: chiral-symmetry breaking by the spin-polarized surface due to the CISS effect and self-amplification by conglomerate crystallization. The symmetry breaking by magnetic surfaces is prebiotically plausible, non-destructive, and robust at room temperature and in solution. Although the demonstrated symmetry-breaking is a surface effect (2D), conglomerate crystallization allows for the extension into the bulk (3D). Therefore, the attained selectivity at the surface level can be seamlessly carried into the bulk of the solution. 

Our mechanism requires a well-defined magnetic surface and the crystallization on colloidal magnetic particles is not enantioselective. A detailed discussion on this can be found in Supplementary Information Section 7.1. In light of this fact, we consider sedimentary rock surfaces similar to those found in the Gale crater \cite{hurowitzetal2017}, as opposed to colloidal muds, as likely sites to realize our mechanism.

With the feedback effect we suggested, our mechanism offers a persistent and deterministic chiral bias to prebiotic chemistry as opposed to a singular and stochastic trigger. Also, by combining the symmetry-breaking with a well-matched amplification mechanism we have shown a direct way to reach the homochiral state from a completely racemic starting point.

The mechanism is a kinetic-entrainment-like effect and the magnetic surface acts as a chiral seed based on its spin polarization direction. Due to the kinetic nature of the effect beyond a certain point enantioselectivity will decrease with increasing crystallization yields. Also, the effect is concentration dependent and it works best around or below the solubility limit of the compound. At higher concentrations, the selectivity goes down as the eutectic equilibrium forces the crystallization. Therefore the slower the crystallization process and the lower the concentration, the higher the enantioselectivity. The low solubility of RAO in water allowed for obtaining its crystals at low concentrations, therefore, we could achieve high enantioselectivity in just one crystallization in comparison to previous work by Tassinari \textit{et al}. with highly soluble amino acids in water \cite{tassinari2019enantioseparation}. Our results can be further improved if the magnetic surface is placed vertically such that the crystals forming in the bulk do not fall on the magnetic surface and reduce the selectivity, as demonstrated by Bhowmick \textit{et al.} \cite{bhowmick2021simultaneous}.

In addition, we used magnetite as the magnetic substrate with near unity spin polarization at its Fermi level \cite{schmitt2021bulk} in comparison to commonly used Ni/Au substrates with lower intrinsic spin polarization (around $20\%$) and protective gold layer further reducing the spin polarization \cite{tassinari2019enantioseparation, bhowmick2021simultaneous}. Therefore, we have shown that magnetite surfaces can be used effectively in the CISS experiments due to their optimal spin-polarization properties.

Our work differs from the previous work by Tassinari \textit{et al}. and Bhowmick \textit{et al}. in multiple ways. First, we achieve much higher ee's due to the low solubility of RAO and the higher spin polarization of magnetite. High ee's under pristine laboratory conditions are necessary to reach homochirality under realistic conditions. Moreover, the crystallization of aminoacids require their supersaturated solutions ($\sim 1$g/mL) in contrast to RAO that is much less soluble ($\sim 0.01$g/mL) in water. Additionally, they achieve the selective crystallization of glutamic acid and threonine only in hydrochloric acid (HCl), asparagine being the exception. This harsh condition is prebiotically implausible and HCl is known to dissolve magnetite, creating iron chlorides. Finally, they show that for a given pole (e.g. north) direction the selective crystallization of asparagine (L) yields the opposite handedness for the cases of glutamic acid (D) and threonine (D). Therefore, the pioneering work by Tassinari \textit{et al}. and Bhowmick \textit{et al}. establish the feasibility of the enantioseperation mechanism using magnetic substrates, however, they do not account for the origin of homochirality.

Although we used stronger magnetic fields to spin-polarize the magnetite surfaces than what was likely available on early Earth, similar spin-polarizations under weak fields can be achieved if magnetite forms as small ($\sim100$ nm in diameter) superparamagnetic particles. Such authigenic magnetite particles have been shown to possess remanent magnetizations of above 10 kA/m even when they form under weak magnetic fields ($0.1$ mT $\approx0.08$ kA/m) comparable to the Earth's \cite{dunlop2005magnetic}. Because our mechanism is based on spin-exchange interaction rather than a magnetic field effect, the figure of merit is not the magnetic field strength but the degree of spin alignment (magnetization) at the active surface. Hence one should pay attention to the prebiotically relevant remanent magnetizations available in natural magnetic minerals rather than early Earth's magnetic field strength. Having said that, with the described feedback mechanism between the chiral molecules and the magnetic surface even higher magnetizations can be realized.

RAO is not just a chiral molecule, it has a central role in the synthesis of ribonucleotides and its chirality directly determines the chirality of RNA. It is very stable against isomerization unlike sugars like ribose or glyceraldehyde as well as against thermal degradation and UV damage. Therefore it is a very suitable molecule to lock the chirality in prebiotic chemistry. Combining this with the fact that it crystallizes as a conglomerate, RAO is an ideal molecule to apply the described enantioseparation mechanism. However, even though we chose RAO as the compound and magnetite ($\ce{Fe3O4}$) as the magnetic surface, our mechanism is more versatile. It is applicable to any chiral molecule forming conglomerate crystals (e.g asparagine, peptides, etc.) and with any prebiotically available magnetic sediment (e.g. maghemite, greigite, etc.). 

\pdfbookmark[section]{\textbf{Acknowledgments}}{sec8}
\section*{Acknowledgments}

The authors thank Oren Ben-Dor, Debkumar Bhowmick, Donna Blackmond, Roger Fu, Stephen Mojzsis, Ron Naaman, and Jack Szostak for helpful discussions, suggestions, and feedback. We thank Victor Loi for the AFM measurements. We acknowledge the Center for Macromolecular Interactions at Harvard Medical School for the use of their CD spectrometer and Dr. Kelly L. Arnett for assistance. We acknowledge the Center for Nanoscale Systems (CNS) and the NSF’s National Nanotechnology Coordinated Infrastructure (NNCI) for the use of the Sharon EE-3 E-Beam Evaporator. We acknowledge the Laukien-Purcell Instrumentation Center for the use of their NMR facility and Dongtao Cui for the assistance. We acknowledge Harvard X-ray Laboratory supported by the Major Research Instrumentation (MRI) Program of the National Science Foundation (NSF) under award number 2216066 and Dr. Shao-Liang Zheng for their help with the X-ray data collection and structure determination. We also acknowledge other members of the Simons Collaboration on the Origins of Life and the Harvard Origins of Life Initiative for fruitful discussions that shaped the ideas behind this work. This work was supported by a grant from the Simons Foundation 290360 to D.D.S.

\section*{Author contributions}

S.F.O., J.D.S., and D.D.S. designed the research; S.F.O. performed the crystallization experiments, collected and analyzed the data; Z.L. synthesized the aminooxazolines; and S.F.O., J.D.S., and D.D.S. wrote the paper.

\section*{Competing interests}

The authors declare no competing interests.

\section*{Data availability}

The data that support the findings of this study are available within the paper and its Supplementary Information

\printbibliography

\appendix
\newpage%

\section{Methods}

We obtained and used the samples without additional purification unless otherwise specified. We performed all of the experiments under ambient conditions unless otherwise specified. \\

\noindent \textbf{Method for Synthesizing RAO, AAO, XAO, and LAO} 

\noindent We synthesized ribo-, arabino-, xylo-, and lyxo-aminooxazolines by the reaction of two equivalents of cyanamide with one equivalent of the corresponding aldopentose sugar. We separately purchased the L and D sugars and synthesized the L and D aminooxazolines in large amounts (above 250 mg each). We then weighed the enantiomer on the scale, ground them into powder, and mixed them in equal amounts to make the racemic pentose aminooxazolines. We confirmed that the aminooxazolines are racemic by CD measurements.\\

\noindent \textbf{Method for Fabricating Magnetite Surfaces}

\noindent We fabricated the magnetite films by evaporating a 100 nm iron layer on 0.625 mm thick silicone (100) wafers using electron beam evaporation under a high vacuum of $5\times10^{-6}$ Torr. Following the evaporation, we baked the samples at 175$^{\circ}$C for 4 hours in the air and promoted the oxidation of iron ($\ce{Fe}$) to magnetite ($\ce{Fe3O4}$). We then cleaned the sample surfaces with acetone and subsequently in ethanol before every experiment. \\ 

\noindent \textbf{Method for Crystallization Experiments} 

\noindent For the direct crystallization experiments we prepared a 0.5M:0.5M solution of racemic glyceraldehyde and 2-aminooxazole in 2 mL water. We incubated the solution at 40$^{\circ}$C for 12 hours and obtained a yellow-brown solution of aminooxazoles. 

For the re-crystallization experiments, we prepared a 65 mM solution of racemic RAO in 2 mL water. 

We placed the magnetite surfaces horizontally in a polystyrene Petri dish (35 by 10 mm) on a magnet such that the surface normal is parallel to the magnetization direction. The magnetic field strength at the sample position was measured to be 325 mT with a Hall probe. We then filled the Petri dish with the incubated solution and made sure that the magnetite surface is covered with the liquid. We then placed the setup in the fridge kept at 12$^{\circ}$C and waited until the first crystals appear. This process can take several hours to a few days. Afterward, we slowly filtered out the mother liquid and washed the surface and crystals with pure water three times such that the racemic liquid is washed away. We then collected the crystals with tweezers under a stereo-microscope. We discarded the crystals formed on the plastic surface of the Petri dish and on the rough edges of the silicon substrate and collected the rest formed on the magnetite surface. We fully dissolved all of the collected crystals in pure water and analyzed the solution.\\

\noindent \textbf{Method for Circular Dichroism Measurements}

\noindent We took the CD measurements in a quartz cuvette after diluting the solution in 2 mL water until the UV/Vis absorption peak is below OD = 1 for accurate measurements. We used a Jasco J-815 Circular Dichroism Spectropolarimeter with an active temperature control connected to a water bath with a temperature set to 20$^{\circ}$C. The temperature feedback is performed by a Jasco PFD-425S/15 controller with a Peltier control unit. Before the measurements, we took a baseline measurement of the water and cuvette background. We simultaneously measured the circular dichroism, and UV/VIS absorption of the sample together with the photo-multiplier voltage to ensure that the spectrometer is not operating beyond its specified voltage range of 600 units. We took the measurements in the 185-210 nm wavelength range and used the auto-baseline subtraction feature. We used a data pitch of 0.2 nm, a bandwidth of 1 nm, a data integration time of 1 second, and a scanning speed of 20 nm/minute. We averaged each measurement 5 times. We normalized the CD signal amplitude based on the UV/Vis absorption and calculated the enantiomeric excess using the calibration procedure described in the Supplementary Information Section 9.

\end{multicols}

\setcounter{figure}{0}    

\begin{figure}[ht]
    \centering
    \renewcommand{\figurename}{Extended Data Fig.}
    \includegraphics[width = 0.8\textwidth]{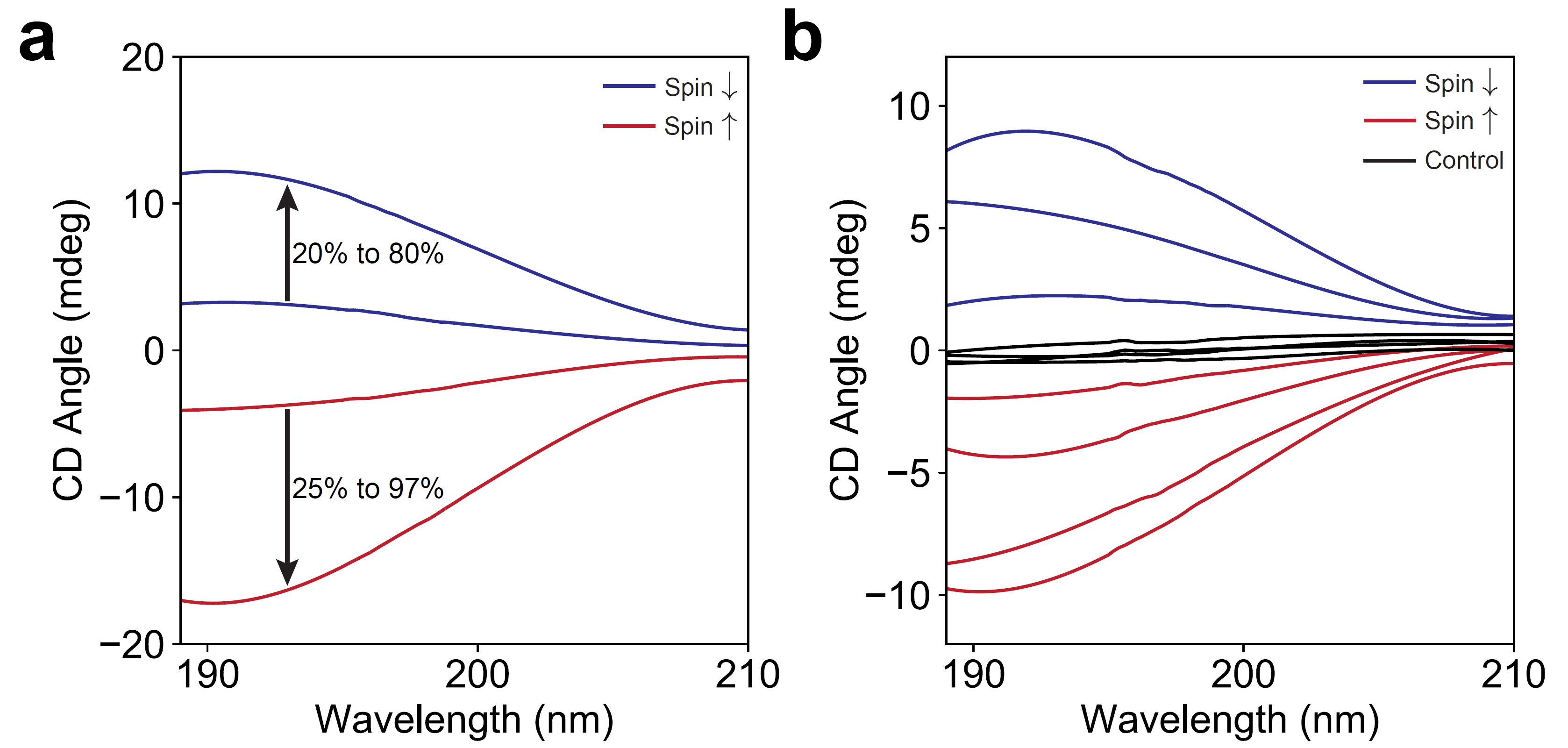}
    \caption{\textbf{Re-crystallization of enantioenriched and racemic RAO on magnetite} \textbf{a.} When we crystallized enantioenriched solutions of RAO we observed a drastic increase in the ee and from about $25\%$ ee we could reach a homochirality state. The $25\%$ enriched D-RAO (red) is crystallized on up-spin magnetite and we obtained homochiral D-RAO. Similarly, $20\%$ enriched L-RAO (blue) is crystallized on down-spin magnetite and we obtained nearly homochiral ($80\%$) L-RAO. \textbf{b.} Repeated re-crystallization of racemic RAO is done to accumulate statistics. DL-RAO is crystallized on up- and down-spin magnetite. Red curves correspond to up-spin and blue curves to down-spin. Black lines are the control experiments done on a silicon surface in the presence of the magnetic field. The ee for the down spin experiments is calculated to be 59, 36, 18 (mean = $38\%$). The ee for the up spin experiments is calculated to be: -10, -22, -41, -54 (mean = $-32\%$). The ee for the control experiments is calculated to be: 5.5, 0.9, 0.6, -3.4 (mean = $0.9\%$). We estimate an error of $\pm5\%$ for the ee calculations.}
    \label{Fig.Ex1}
\end{figure}

\begin{figure}[H]
    \centering
    \renewcommand{\figurename}{Extended Data Fig.}
    \includegraphics[width = 0.7\textwidth]{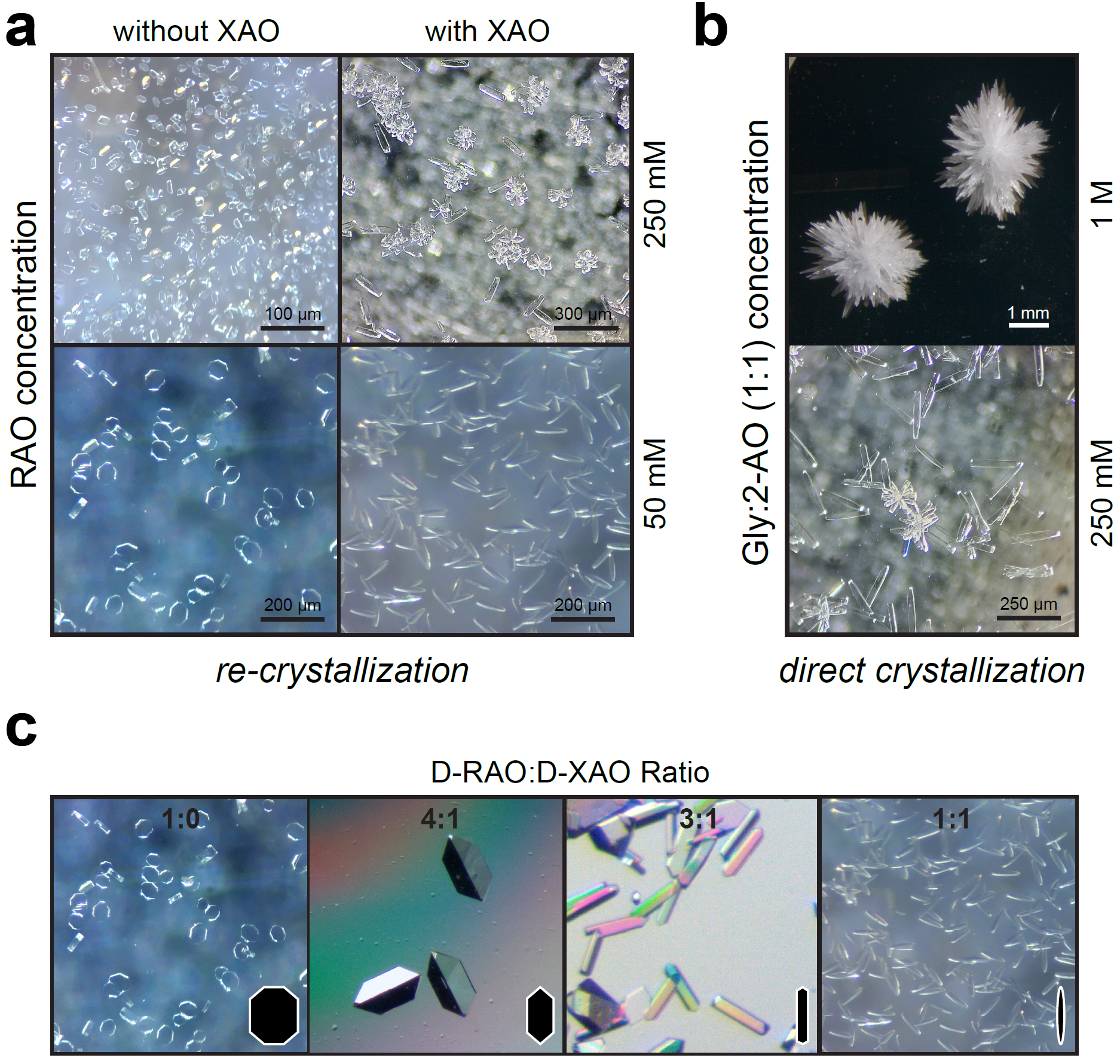}
    \caption{\textbf{Crystal habit modification of RAO by XAO} \textbf{a.} RAO crystallization is modified with XAO. In the presence of XAO and at high concentrations, RAO forms rosette-shaped twinned conglomerates. \textbf{b.} Direct crystallization of RAO gives rosette-shaped twinned clusters, and as the concentration is decreased the conglomerate crystals become more angular and the rosette formation is suppressed. At lower concentrations for direct crystallization, the crystal morphology becomes identical to re-crystallized RAO in the presence of XAO. \textbf{c.} XAO modifies the crystal shape of RAO by inhibiting the growth of the side face and thus promoting the growth vertically. This makes the RAO crystals more elongated and eventually needle-shaped with increasing XAO abundance. On the top-center, the ratio between D-RAO and D-XAO; on the bottom-right, the sketch of the crystal shape is shown.}
    \label{Fig.Ex2}
\end{figure}

\end{document}